\documentclass[10pt]{article}
\usepackage{cite,epsf,graphicx,color,axodraw,subfigure,psfrag}

\definecolor{darkmagenta}{rgb}{.45,0,.45}
\def\lsim{\mathrel{\rlap{\lower4pt\hbox{\hskip1pt$\sim$}}
    \raise1pt\hbox{$<$}}}                
\def\gsim{\mathrel{\rlap{\lower4pt\hbox{\hskip1pt$\sim$}}
    \raise1pt\hbox{$>$}}}                

\newcommand{\gev}{\,{\rm GeV}}
\newcommand{\mxsq}{{M_X^2}}
\newcommand{\qsq}{{Q^2}}
\newcommand{\pt}{{p_{\bot}}}
\newcommand{\ptsq}{{p_{\bot}^2}}
\newcommand{\xpom}{{x_{I\!\!P}}}
\newcommand{\pom}{{I\!\!P}}
\newcommand{\rmd}{{{\rm d}}}
\newcommand{\ptsqmax}{{p_{\bot\,{\rm max}}^2}}
\newcommand{\ptsqmin}{{p_{\bot\,{\rm min}}^2}}
\newcommand{\intercept}{{\alpha_{I\!\!P}(0)}}
\begin{document}

\title{\hfill{\small OUTP-99-56P} \\
\bigskip
Transverse Momentum Structure of Diffractive DIS Models}
\author{J.C. Williams\thanks{email: j.williams1@physics.ox.ac.uk}\\
Theoretical Physics,
        Oxford University, \\
        1 Keble Road, \\
        Oxford, OX1 3NP, U.K.%
}
\date{\today}
\maketitle

\begin{abstract}
The transverse momentum distribution of the diffractive final state
provides an interesting test of models of diffractive deep-inelastic
scattering at HERA. We present a comparison of several colour-singlet
exchange models with thrust transverse momentum data from a recent H1
analysis. We also study the effect of constraints imposed on the
kinematically-accessible phase space by data selection cuts and find
that, as a result of the pseudo-rapidity cut which is used by H1 to
select diffractive events, there is no dijet contribution at low
transverse momenta. We are able to fit the large transverse momentum
part of the data with a two-gluon dijet model. The results of this
analysis are compared with a previous study of large pseudo-rapidity
gap structure function data, and we discuss ways in which one might
reconcile the results of the two analyses. We conclude that a
significant small-$\beta$ 3-jet contribution is probably required to
explain the data, and show that the combination of a two-gluon dijet
model and an exponentially-decaying thrust transverse momentum
distribution provides a good fit over the entire kinematic range of
the thrust data.
\end{abstract}


\section{Introduction}

With the increasing numbers of
models\,\cite{Ellis:1996cg,Donnachie:1987pu,Donnachie:1987xh,Vermaseren:1996iy,Landshoff:1987yj,Diehl:1995wz,Diehl:1998pd,Bartels:1998ea}
which propose to describe diffractive deep-inelastic scattering (DIS),
it is important to find ways to test them. One method is to examine
events with a large pseudo-rapidity gap between the proton remnant and
the diffractive final state. The requirement of a large
pseudo-rapidity gap means that the experimental cuts are selecting a
sample with a strongly reduced phase space\,\cite{Ellis:1996cg}. An
analysis of the effects of pseudo-rapidity cuts in terms of
diffractive structure functions has been reported
previously\,\cite{Ellis:1998qt}. A further test of models of
diffractive DIS is to study the unintegrated transverse momentum
distribution of diffractive
events\,\cite{Donnachie:1992rh,Buchmuller:1997xw,Buchmuller:1997eb}.

Here we provide a detailed analysis of the thrust transverse momentum
distribution reported recently by H1\,\cite{Adloff:1997nn}. We study
the restrictions on available phase space for diffractive interactions
which are imposed by a combination of pseudo-rapidity cuts and other
data selection cuts. We find that these cuts reject dijet
(quark--antiquark diffractive final state) events with small
transverse momenta relative to the virtual photon--proton axis. An
additional contribution is therefore required to fit the data at small
transverse momenta. At the values at which the low-momentum dijet
cut-off occurs the data also changes slope, having an approximately
quadratic fall-off with increasing transverse momentum up to about
4-10$\gev^2$ which changes to a quartic fall-off at larger transverse
momenta. The ($\beta$-dependent) position of the low transverse
momentum dijet cut-off corresponds rather closely to the transverse
momentum values at which the slope of the cross section changes.

In this paper we consider how dijet events contribute to the
transverse momentum distribution, and discuss the r\^ole that
higher-multiplicity diffractive final-state events are expected to
play. The colour-singlet exchange models of dijet production which we
consider here are: the form factor vector pomeron models of Donnachie
and Landshoff (D-L)\,\cite{Donnachie:1987pu,Donnachie:1987xh} and
Ellis and Ross (E-R)\,\cite{Ellis:1996cg,Ellis:1998qt} (in the Feynman
gauge and in a non-covariant gauge), the scalar pomeron model of
Vermaseren {\em et al}\,\cite{Vermaseren:1996iy}, a leading-twist
single-gluon exchange model, and the two-gluon Landshoff-Nachtmann
model described in\,\cite{Diehl:1995wz}. The Donnachie-Landshoff
(non-covariant gauge) and two-gluon models provide a good fit to the
quartic fall-off of the diffractive cross section with thrust
transverse momentum, while the other models predict only a quadratic
fall at large transverse momenta.

The result of the thrust transverse momentum analysis appears, at
first, rather in contradiction with a recent analysis of large
pseudo-rapidity gap structure function data\,\cite{Ellis:1998qt}. The
structure function analysis appeared to strongly favour the $1/\ptsq$
models. In particular, the E-R and leading-twist models fit the
structure function data well over a wide range of $\qsq$, $\beta$ and
$\xpom$. The two-gluon and Donnachie-Landshoff models fall off too
rapidly with decreasing $\beta$ to fit the data without a substantial
contribution from higher-order diffractive final states. We explore
two possible ways to reconcile these results: the first is to consider
that the models which give a quadratic dijet transverse momentum
dependence describe the diffractive data only at large $\beta$, while
the small-$\beta$ spectrum is dominated by a two-gluon-like
contribution. This is a possibility that cannot be ruled out by
present large pseudo-rapidity gap data. The other explanation for
these results is that both sets of data might be described by the
combination of a two-gluon dijet component and a (possibly
non-perturbative) contribution which is strongly peaked at small
$\beta$ and small thrust transverse momentum.

The position of the low transverse momentum cut-off is largest at
small values of $\beta$. This means that the extra thrust transverse
momentum contribution which is required to ``fill in'' the thrust
transverse momentum distribution spectrum is greatest at large
diffractive masses, i.e., at small $\beta$. Since $\beta$ plays the
r\^{o}le of the scaling variable $x$ in standard DIS, we conclude that
the additional contribution is probably due to the gluonic structure
of the pomeron. That is, the additional contribution needed to
describe the thrust transverse momentum spectrum at small transverse
momenta is probably due to 3-or-more parton diffractive final states,
predominantly quark--antiquark--gluon (``3-jet'') diffractive final
states. Another reason for this conclusion is that most of the H1
thrust transverse momentum distribution data corresponds to very small
values of $\beta$, and it is at such such values that one might expect
the contribution from 3-jet final states to be
significant\,\cite{Wusthoff:1997fz,Bartels:1998ea,Bartels:1999tn}. The
combination of a two-gluon dijet model and a simple $\beta$-dependent
Gaussian fits the data well over the entire kinematic range in the H1
analysis.

The rest of this paper is organized as follows: in
Sect.\,\ref{sect:bookkeeping} we present a discussion of the usual
diffractive DIS kinematics, in which we introduce the kinematic
variables which are of particular interest in this
analysis. Sect.\,\ref{sect:ptsqcuts} contains an analysis of the
effects of data selection cuts. In Sect.\,\ref{sect:ptsqanalysis} we
describe several models of diffractive dijet production and compare
these models with thrust transverse momentum data from a recent H1
study\,\cite{Adloff:1997nn}. In Sect.\,\ref{sect:3jets} we compare our
results with a recent analysis of large pseudo-rapidity gap structure
function data and discuss two possible ways to reconcile the two
analyses. This section includes a discussion of the possible
contribution of 3-jet diffractive final states to the transverse
momentum spectrum. Sect.\,\ref{sect:conclusions} summarises the
results presented in this paper and outlines further investigations
which could be carried out to verify our conclusions. In the Appendix
we give a brief discussion of the large pseudo-rapidity gap structure
function analysis carried out previously\,\cite{Ellis:1998qt}. In this
Appendix we extend the comparison of form factor models with large
pseudo-rapidity gap structure function data to allow for a different
form factor cut-off, $\Lambda$, and include fits for the form factor
models calculated in the non-covariant gauge.

\section{Background} \label{sect:bookkeeping}

\subsection{Kinematics} \label{sect:kinematics}

\subsubsection{Standard Kinematics}

In the HERA electron--proton experiments 820\,GeV
protons\footnote{920\,GeV protons since August 1998.} collide with
27.5\,GeV electrons or positrons. This corresponds to a centre-of-mass
(CMS) energy of $\sqrt{S_{\rm Tot}}\sim 300\gev$. In the HERA lab
frame the positive~$z$-axis is defined by the forward proton direction
with the origin at the interaction vertex.  We consider diffractive
deep-inelastic $e-P$ scattering,

\begin{equation}
e(p_{e}) + P(P) \rightarrow e(p_{e}^{\prime}) + P(P^{\prime}) + X(X),
\end{equation}

\noindent
where the momenta of the particles are shown in brackets. One may
consider that the interaction proceeds by virtual photon--pomeron
deep-inelastic scattering,

\begin{equation}
\gamma^{*}(q) + \pom(P_{\pom}) \rightarrow X(X),
\end{equation}

\noindent where~$P_{I\!\!P}=P-P^{\prime}$.

We use the usual kinematic variables of deep-inelastic scattering,

\begin{equation} \label{eq:standardkinematics}
Q^2=-q^2,~~~x=\frac{Q^2}{2P\cdot q},~~~{\rm
and}~~y=\frac{Q^2}{x\,S_{\rm Tot}},
\end{equation}

\noindent where~$Q^2$ is the negative four-momentum squared of the virtual
photon and~$x$ is the Bjorken scaling variable. We also define~$W^2$, the
mass squared of the total hadronic system ($X\,+\,$outgoing proton), by

\begin{equation}
W^2=(P+q)^2.
\end{equation}

Additionally, for diffractive scattering we define:

\begin{equation}
t_{I\!\!P}=(P-P^{\prime})^2,~~~x_{I\!\!P}=\frac{Q^2+M_{X}^2}{Q^2+W^2},
~~~{\rm and}~~\beta=\frac{Q^2}{Q^2+M_{X}^2},
\end{equation}

\noindent
where~$t_{I\!\!P}$ is the momentum transfer at the proton vertex and
is constrained by experimental cuts to be small ($%
|t_{I\!\!P}|\lsim 1$\,GeV$^2$),~$x_{I\!\!P}$ is the fraction of
longitudinal momentum of the proton carried by the pomeron,
and~$x=\beta x_{I\!\!P}$. The mass squared of the diffractive
system~$X$ is~$M_{X}^2$, and the proton mass is neglected.

The {\em pseudo-rapidity},~$\eta$, of an outgoing particle is defined
in the laboratory frame in terms of its polar angle with respect to
the proton direction by

\begin{equation}
\eta=-\ln\tan\left(\frac{\theta_{\rm lab}}{2}\right).
\end{equation}

\noindent

\subsubsection{Exchanged Quark Virtuality, Thrust and Transverse
Momentum}\label{sect:transversemomentum}

\begin{figure}[htp]
\begin{center} \begin{picture}(180,130)(0,0)
\SetScale{0.6}
\Line(5,128)(60,128)
\Line(60,128)(105,155)
\Line(84,103)(130,103)
\Line(84,58)(130,58)
\Line(84,103)(84,58)
\Line(30,13)(130,13)
\Photon(60,128)(84,103){4}{3.5}
\Text(60,22)[]{\scriptsize $\pom$}
\Text(-1,76)[]{\scriptsize $e$}
\Text(68,95)[]{\scriptsize $e$}
\Text(83,62)[]{\scriptsize $\bar{q}$}
\Text(83,35)[]{\scriptsize $q$}
\Text(45,48)[]{\scriptsize $k$}
\Text(12,9)[]{\scriptsize $P$}
\Text(83,9)[]{\scriptsize $P$}
\SetColor{Blue}
\ZigZag(84,58)(84,13){-4}{4.5}
\SetColor{Black}
\Vertex(60,128){1}
\Vertex(84,103){1}
\Vertex(84,58){1}
\Vertex(84,13){1}
\end{picture}
\caption[Diffractive DIS via pomeron exchange.]{{\em Diffractive DIS
via pomeron exchange.}\label{fig:dijet}} 
\end{center}
\end{figure}
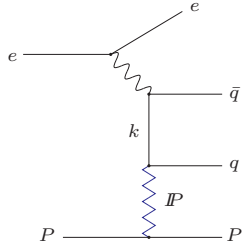

For the leading-order (to order $\alpha_s$) diffractive process shown
in~Fig.\,\ref{fig:dijet} we introduce a further invariant, the
four-momentum squared of the struck quark,~$k^2$. In
the~$\gamma^{*}-I\!\!P$ CMS the virtuality of this quark can be
expressed in terms of other invariants and the polar angle with
respect to the~$\gamma^{*}-I\!\!P$ axis, by

\begin{equation}
k^2= -\frac{Q^2+M_{X}^2}{2}(1-\cos\theta_{\rm cms}).  \label{eq:ksqdefn}
\end{equation}

\noindent
A similar expression can be formed for interactions where more than
two diffractive final-state partons are produced.

There are two further quantities of interest here. The first is {\em
thrust}, $T$, which describes the principal axis of jet momentum and
the spread of final-state particles. More formally, the thrust axis is
defined to be parallel to the vector $\vec{n}$ which maximises the
weighted sum

\begin{equation} \label{eq:thrustdefinition}
T=(\frac{1}{\sum_{i=1}^{N}|\vec{p}_i|})\cdot\max_{\vec{n}}\sum_{i=1}^{N}|\vec{p}_i\cdot\vec{n}|,
\end{equation}

\noindent
where the sum is taken over the whole diffractive final state and $T$
is the thrust variable\footnote{Some references use a slightly
different definition of $T$. The normalisation of $T$ we have adopted
here is that used in\,\cite{Adloff:1997nn}.}. The thrust axis in
diffractive dijet production is the axis defined by the back-to-back
final-state quarks in the virtual photon--pomeron CMS and is
approximated experimentally by the thrust axis described by the
(hadronic) diffractive final state. For a dijet event $T$ is equal to
1, while for a symmetric 3-jet event $T=2/3$ and a completely
isotropic multi-jet event will have $T=1/2$. In general, $T$ can take
any value between 1/2 and 1.

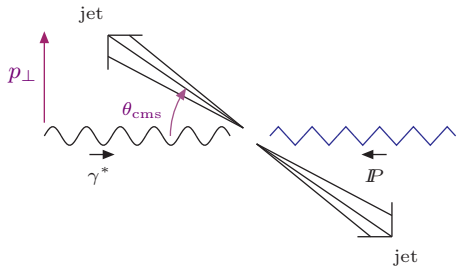
\begin{figure}[htp]
\begin{center} \begin{picture}(256.5,96)(0,0)
\SetScale{1}
\LongArrow(62,43)(70,43)
\Photon(45,50)(115,50){3.5}{5.5}
\LongArrow(174,43)(166,43)
\SetColor{Blue}
\ZigZag(130,50)(200,50){3.5}{5.5}
\SetColor{Black}
\Text(66,35)[]{\scriptsize $\gamma^*$}
\Text(170,35)[]{\scriptsize $\pom$}
\Line(120,53)(69,88)
\Line(120,53)(69,80)
\Line(120,53)(77,88)
\Line(69,88)(69,75)
\Line(69,88)(82,88)
\Line(125,47)(176,12)
\Line(125,47)(176,20)
\Line(125,47)(168,12)
\Line(176,12)(176,25)
\Line(176,12)(163,12)
\Text(63,96)[]{\scriptsize jet}
\Text(182,4)[]{\scriptsize jet}
\SetColor{RedViolet}
\LongArrow(45,55)(45,88)
\Text(37,73)[]{\small\textcolor{darkmagenta}{$p_{\bot}$}}
\SetColor{Black}
\SetColor{DarkOrchid}
\LongArrowArcn(122.5,50)(30,180,143)
\SetColor{Black}
\Text(90,60)[r]{\scriptsize\textcolor{darkmagenta}{$\theta_{\rm cms}$}}
\end{picture}
\caption[Definition of transverse momentum variable $\ptsq$ (in
$\gamma^*-\pom$ CMS).]{{\em Definition of transverse momentum variable
$\ptsq$ (in $\gamma^*-\pom$ CMS).}}\label{fig:ptsqfigure}
\end{center}
\end{figure}

Another useful kinematic quantity is the thrust transverse momentum,
$\ptsq$, defined in the virtual photon--pomeron CMS. For events with
dijet final states, $p_{\bot}$ is simply the component of momentum of
one of the final state hadronic jets which is transverse to
photon--proton axis (equivalently, to the photon--pomeron
axis\footnote{Here we assume that, for small $t_{\pom}$, the pomeron
is emitted parallel to the proton direction.}). This is shown in
Fig.\,\ref{fig:ptsqfigure}. In the more general case of final states
with two or more partons, the experimentally measured transverse
momentum variable is the transverse momentum of the thrust axis with
respect to the photon--proton direction.

The transverse momentum variable defined above is different to that
defined in\,\cite{Williams:1999ze} where phase space constraints due
to pseudo-rapidity cuts\,\cite{Ellis:1996cg,Ellis:1998qt} are
discussed in terms of a lower limit on the transverse momentum of the
final-state parton which couples to the pomeron. The effect of
pseudo-rapidity cuts on the 3-jet thrust transverse momentum
distribution is more subtle as there exist 3-jet final-state
configurations in which $\ptsq_{\rm thrust}=0$, for example, but for
which the events will survive most pseudo-rapidity cuts. Events where
there is a parton emitted far in the forward direction, however, are
expected to be rejected by pseudo-rapidity cuts and hence there is
likely to be a systematic reduction in the low-$\pt$ 3-jet thrust
transverse momentum spectrum. Contraints on thrust transverse momentum
due to data selection cuts could, at best, be investigated through
Monte Carlo studies of diffractive DIS.

\section{Phase Space Analysis of Data Selection Cuts} \label{sect:ptsqcuts}
 
We start this section with a summary of the data under investigation
and then discuss the effects of the cuts used to select the
diffractive DIS sample.

H1 measured the diffractive DIS cross section as a function of thrust
transverse momentum, $\ptsq$, and final-state diffractive mass, $M_X$.
They presented their results in the form\,\cite{Adloff:1997nn}

\begin{equation}
\left.\frac{1}{\sigma}\,\frac{\rmd^2\sigma}{\rmd\mxsq\,\rmd\ptsq}
(\ptsq)\frac{}{}\right|_{\,{\rm fixed}~M_X}.
\end{equation}

\noindent
The diffractive cross section has been integrated over $\qsq$ and $\xpom$ in the
range

\begin{eqnarray}
\qsq &:&10\to 100\,{\rm GeV}^2\,, \nonumber \\
\xpom &:&10^{-4}\to 3.2\times 10^{-2},
\end{eqnarray}

\noindent
and the data was binned with average diffractive masses of

\begin{equation}
M_X= 7.01,~9.40,~12.82,~16.85,~21.20,~28.58\,\gev.
\end{equation}

\noindent
This corresponds to integrating over small values of $\beta$, for
example: for $M_X=7.01\gev$, over the range of $\qsq$ in this analysis
$\beta$ ranges from 0.17 to 0.67; whereas for $M_X=28.58\gev$, $\beta$
ranges from 0.01 to 0.11. Here ``small'' is to be interpreted as
meaning values of $\beta$ for which one might expect 3-jet processes
to provide a significant contribution to the diffractive final
state (specifically, $\beta\lsim0.3$\,\cite{Bartels:1998ea}).

The H1 transverse momentum distribution data\,\cite{Adloff:1997nn} has
a $1/\ptsq$ fall-off at low transverse momenta, steepening to a
$1/p_{\bot}^4$ fall-off at higher transverse momenta. This change in
slope occurs between about 1-10$\gev^2$ for the values of $M_X$ in
this study.  The $1/p_{\bot}^4$ tail in the data is not a phase space
effect, as this tail is present for all values of $M_X$ a long way
away from the phase space limit of $\ptsqmax=\mxsq/4$. For
$M_X=28.58\gev$, for example, the data changes slope at about
$\ptsq=10\gev^2$.

In an analysis of this data one must take account of the effects of
experimental cuts on the space of $\qsq$ and $\xpom$ which was
integrated over. There are two cuts which restrict the accessible
range of $\qsq$ and $\xpom$ for a given final-state transverse
momentum and diffractive mass:

\begin{itemize}
\item[1.] {\it Cut on $y$: $y<0.5$.}
\end{itemize}

This cut is imposed to restrict the photo-production background, as
well as removing events where the diffractive system is strongly
boosted in the backward direction\,\cite{Adloff:1997nn}. Using the
relation\,(\ref{eq:standardkinematics}) between $y$ and the other
kinematic variables we get the restriction on $\xpom$,

\begin{equation}
\xpom>\frac{2(\qsq+\mxsq)}{S_{\rm Tot}},
\end{equation}

\noindent
which corresponds to a lower limit on the integral over $\xpom$ as a
function of $\qsq$ at fixed $M_X$.

\begin{itemize}
\item[2.] {\it Pseudo-rapidity cut: require pseudo-rapidity gap
greater than $\eta_{\rm max}=3.2$.}
\end{itemize}

It was shown in\,\cite{Ellis:1998qt,Williams:1999ze} that imposing a
pseudo-rapidity cut removes part of the low-$\pt$ dijet contribution
to diffractive DIS. For a given range of kinematic variables $\qsq$,
$\mxsq$ and $\xpom$, a pseudo-rapidity cut corresponds to imposing a
lower transverse momentum cut-off, $\ptsqmin$, on accepted
events. Therefore, when calculating the cross section at a given
$\ptsq$, say $\ptsq_{\rm fixed}$, one should only integrate over the
region of $\qsq$ and $\xpom$ for which $\ptsqmin<\ptsq_{\rm
fixed}$. The region of phase space for which $\ptsqmin>\ptsq_{\rm
fixed}$ will have been removed by the pseudo-rapidity cut.

\begin{figure}[htp]  
\centering \includegraphics[totalheight=0.25\textheight,
]{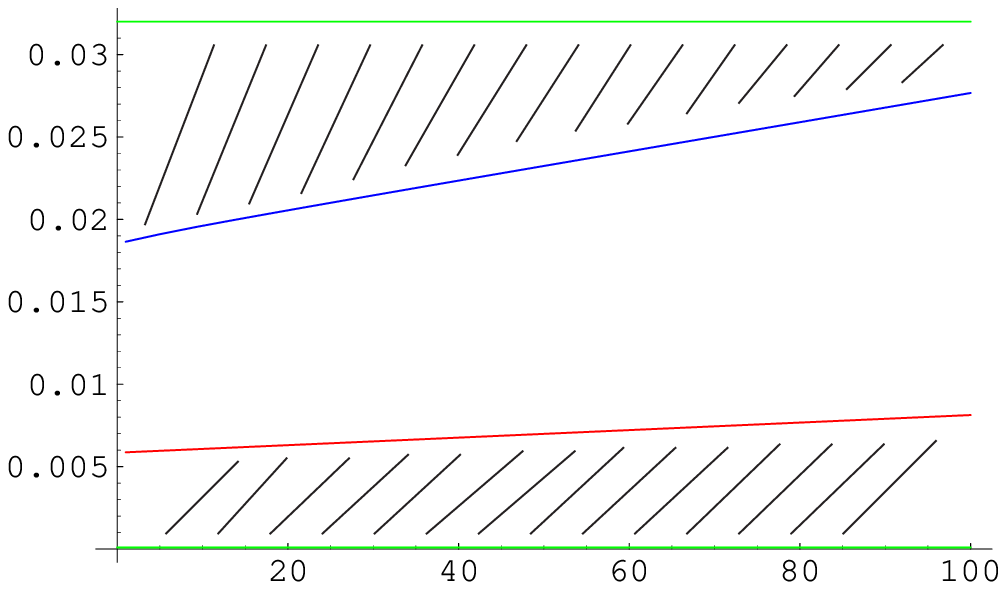}
\begin{picture}(10,10)
\Text(23,5)[]{$Q^2\,\gev^2$}
\Text(-230,140)[]{$\xpom$}
\CBoxc(-125,115)(60,19){White}{White}
\Text(-125,115)[]{$\ptsqmin>4$}
\CBoxc(-95,22)(45,19){White}{White}
\Text(-95,22)[]{$y>0.5$}
\end{picture}
\caption[Minimax plot.]{{\em Phase space plot for $\ptsqmin=4$,
$\eta_{\rm max}=3.2$ (hadron-level cut), and $M_X=16$. The horizontal lines
show upper and lower limits of $\xpom$ in data sample, while the upper
slanted line shows the upper limit on $\xpom$ as a result of
pseudo-rapidity cuts, and the slanted lower line shows the lower limit
of $\xpom$ as a result of the $y$ cut.}}\label{fig:minimaxplot}
\end{figure}

\vspace*{0.3cm}The above discussion of the effect of pseudo-rapidity
cuts on available phase space is illustrated in
Fig.\,\ref{fig:minimaxplot}. In this figure, the region in the
$\qsq-\xpom$ plane over which one would integrate to extract the cross
section at $\ptsq=4\gev^2$ is shown for a diffractive mass of
$M_X=16\gev$ and a pseudo-rapidity cut of 3.2 (typical HERA
values). The upper shaded area shows the region excluded by the
pseudo-rapidity cut. In this region of $\qsq$ and $\xpom$ dijet events
with $\ptsq<4\gev^2$ are removed by the pseudo-rapidity cut. The lower
shaded area shows the region excluded by the cut on $y$. As a result,
there is a restricted region of phase space populated by dijet events
with thrust transverse momentum of 4$\gev^2$.

\begin{figure}[htp]
\centering
		\psfrag{ylabel}{$\frac{1}{\sigma}\frac{\rmd\sigma}{\rmd\ptsq}\gev^2$}
		\psfrag{xlabel}{\small{$\ptsq\gev^2$}}
		\includegraphics[width=.65\textwidth]{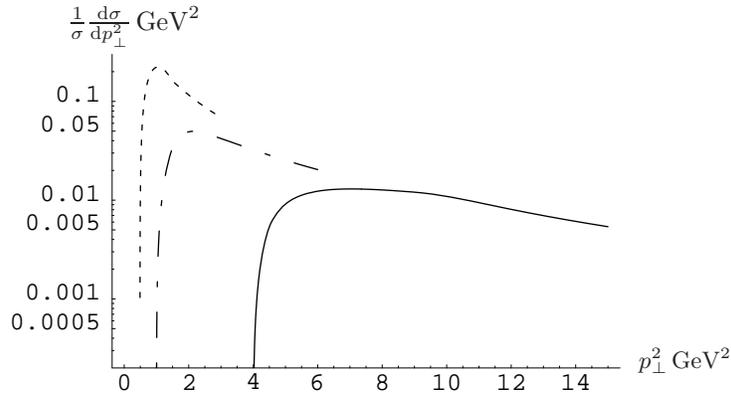}
\caption[ptcutoffs.]{{\em Position of dijet transverse momentum
cut-offs at different diffractive masses (two-gluon model). The solid
line corresponds to a diffractive mass of 28.58\,GeV, while the
dot-dashed line is for 21.20\,GeV, and the dotted line represents a
diffractive mass of 16.85\,GeV.}}\label{fig:ptcutoffs}
\end{figure}

For large diffractive masses and small $\ptsq$, the data selection
cuts exclude the entire $\qsq-\xpom$ plane, leading to a sharp
low-momentum cut-off in $\ptsq$ below which there are no values of the
kinematic parameters $\qsq$ and $\xpom$ in the H1 study for which
dijet events can contribute to the diffractive cross section. The
position of the cut-off depends on $M_X$, and moves to larger
transverse momentum values for larger diffractive masses. In
Fig.\,\ref{fig:ptcutoffs}, the position of the low-momentum dijet
cut-off is shown for three diffractive masses: $M_X=28.58\gev$,
$21.20\gev$, and $16.85\gev$.  For smaller diffractive masses the
transverse momentum distribution of dijet events is not strongly
restricted by the data selection cuts and, for $M_X\lsim15\gev$, there
is no lower $\pt$ cut-off from the data selection cuts.

This analysis of the effect of data selection cuts on the dijet
contribution to diffractive DIS therefore demonstrates that an
additional contribution is needed to describe the H1 thrust transverse
momentum distribution data, particularly at low transverse momenta.

\section{Analysis of Diffractive Dijet Production Models}
\label{sect:ptsqanalysis}

In this section we present the fits of various dijet models to the H1
thrust transverse momentum distribution data, starting with a brief
description of the diffractive dijet models under consideration. The
contribution of higher-order diffractive processes, such as
quark--antiquark--gluon diffractive final states, is discussed in the
following section.

\subsection{Models of Diffractive Dijet Production}
\label{sect:models}

The first type of model we studied is the form factor approach of
Donnachie and Landshoff\,\cite{Donnachie:1987pu,Donnachie:1987xh} and
Ellis and Ross\,\cite{Ellis:1996cg,Ellis:1998qt} in which the
diffractive interaction is assumed to proceed by the exchange of a
$C-$even, vector-like pomeron. This approach allows for a direct
coupling between the pomeron and off-shell partons. In order to
maintain the experimentally-observed scaling of the diffractive
structure function, a {\em form factor} is inserted to soften the
virtual quark--pomeron vertex. In the original phenomenological model
of Donnachie and Landshoff\,\cite{Donnachie:1987pu,Donnachie:1987xh},
applied to deep-inelastic scattering, one inserts a form factor at the
virtual quark--pomeron vertex of

\begin{equation} \label{eq:DLformfactor}
f(k^2)=\frac{\Lambda^2}{\Lambda^2-k^2}.
\end{equation}

\noindent
Here $k^2$ is the four-momentum squared of the exchanged quark
coupling to the pomeron (the other quark at this vertex is near mass
shell and hence its momentum may be neglected), and $\Lambda^2$ is of
the order of the hadronic scale ($\Lambda^2\sim 1\,{\rm GeV}^2$). This
serves to cut off contributions at large parton virtualities. In
the modification suggested by Ellis and Ross\,\cite{Ellis:1996cg}, one
chooses the form factor

\begin{equation} \label{eq:ERformfactor}
f(k^2)=\sqrt{\frac{\Lambda^2}{\Lambda^2-k^2}}.
\end{equation}

\noindent
This choice of form factor leads to a cross section with a logarithmic
dependence on $k^2$, which therefore receives contributions from the
entire $k^2$ spectrum\footnote{Both form factor models reduce to the
original Donnachie-Landshoff model of hadron-hadron
scattering\,\cite{Donnachie:1984hf,Donnachie:1984xq,Donnachie:1986iz,Donnachie:1992ny}
for the exchange of a pomeron between two on-shell partons.}.  This
model is motivated by the observation that diffractive DIS is relatively
insensitive to
pseudo-rapidity\,\cite{Ahmed:1994nw,Ahmed:1995ui,Derrick:1993xh}, as
well as the correlation between between exchanged quark virtuality and
pseudo-rapidity which was demonstrated in\,\cite{Ellis:1998qt}.

The other models we consider were described in detail
in\,\cite{Ellis:1998qt}. The first is a model in which the ``pomeron''
is described by the colour-singlet part of single-gluon exchange, that
is, the exchange of one hard gluon with subsequent exchanges of soft
gluons which ``dress'' the gluon colour enabling the formation of a
pseudo-rapidity gap. We also consider a direct-coupling scalar pomeron
model\,\cite{Vermaseren:1996iy}. Finally, we consider the two-gluon
Landshoff-Nachtmann model\,\cite{Landshoff:1987yj} following the
analysis of Diehl\,\cite{Diehl:1995wz}. In this approach the
colour-singlet state is modeled by the exchange of two non-interacting
gluons which are described by non-perturbative propagators and which
are allowed to couple to the two quarks in all possible combinations.

\begin{figure}[htp]
\begin{center} \begin{picture}(300,95)(40,0)
\SetScale{0.5}
\Line(5,128)(60,128)
\Line(60,128)(105,155)
\Line(84,103)(130,103)
\Line(84,58)(130,58)
\Line(84,103)(84,58)
\Line(30,13)(130,13)
\Photon(60,128)(84,103){4}{3.5}
\Gluon(92,58)(92,13){-4}{4.5}
\Gluon(106,58)(106,13){-4}{4.5}
\Vertex(60,128){1}
\Vertex(84,103){1}
\Vertex(92,58){1}
\Vertex(106,58){1}
\Vertex(92,13){1}
\Vertex(106,13){1}
\Line(175,128)(230,128)
\Line(230,128)(275,155)
\Line(254,103)(300,103)
\Line(254,58)(300,58)
\Line(254,103)(254,58)
\Line(200,13)(300,13)
\Photon(230,128)(254,103){4}{3.5}
\Gluon(262,103)(262,13){-4}{8.5}
\Gluon(276,103)(276,13){-4}{8.5}
\Vertex(230,128){1}
\Vertex(254,103){1}
\Vertex(262,13){1}
\Vertex(276,13){1}
\Vertex(262,103){1}
\Vertex(276,103){1}
\Line(345,128)(400,128)
\Line(400,128)(445,155)
\Line(424,103)(470,103)
\Line(424,58)(470,58)
\Line(424,103)(424,58)
\Line(370,13)(470,13)
\Photon(400,128)(424,103){4}{3.5}
\Gluon(432,58)(432,13){-4}{4.5}
\Gluon(446,103)(446,13){-4}{8.5}
\Vertex(400,128){1}
\Vertex(426,103){1}
\Vertex(432,58){1}
\Vertex(446,103){1}
\Vertex(432,13){1}
\Vertex(446,13){1}
\Line(515,128)(570,128)
\Line(570,128)(615,155)
\Line(594,103)(640,103)
\Line(594,58)(640,58)
\Line(594,103)(594,58)
\Line(540,13)(640,13)
\Photon(570,128)(594,103){4}{3.5}
\Gluon(602,103)(602,13){-4}{8.5}
\Gluon(616,58)(616,13){-4}{4.5}
\Vertex(570,128){1}
\Vertex(594,103){1}
\Vertex(602,103){1}
\Vertex(616,58){1}
\Vertex(602,13){1}
\Vertex(616,13){1}
\end{picture}
\caption{{\em Two-gluon exchange graphs used to model pomeron
exchange.}\label{fig:coupleofgluons}}
\end{center}
\end{figure}
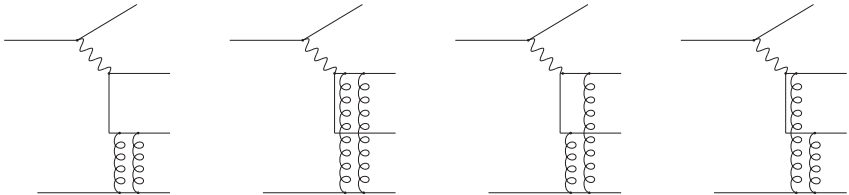

It should be noted that the form factor models are not
gauge-invariant\,\cite{Diehl:1995wz,Diehl:1998pd}. The sum of the
diagrams which contribute to two-gluon exchange, as shown in
Fig.\,\ref{fig:coupleofgluons}, is an explicitly gauge-invariant
quantity. Comparing the two approaches, the diagrams where the
colour-singlet exchange is modeled by a hadron-like ``pomeron''
exclude the two diagrams in which the two gluons do not couple to the
same quark line. At best, therefore, one might hope for the form
factor models to be physically meaningful in a gauge where the
contribution from these two diagrams is minimal.
Diehl\,\cite{Diehl:1998pd} has suggested a non-covariant gauge in
which the D-L model mimics many of the features of the manifestly
gauge-invariant Landshoff-Nachtmann two-gluon model. We find that
studying the unintegrated transverse momentum distribution highlights
the gauge non-invariance problem with the form factor approach, and we
have chosen to calculate the form factor models in the non-covariant
gauge in this analysis.  We have also repeated the structure function
analysis of\,\cite{Ellis:1998qt} for the form factor models in this
gauge (see Appendix).

\subsection{Transverse Momentum Distribution Fits with Dijet Models}

The diffractive scattering cross section is calculated by integrating
the cross section from the colour-singlet exchange models over the
region of phase space in $\qsq$ and $\xpom$ accessed in the H1
analysis. The common feature of all the fits is that, as a result of
the large pseudo-rapidity gap requirement and the cut on~$y$, the
dijet contribution is cut off sharply at low $\ptsq$. For example, for
a diffractive mass of $M_X=28.58\gev$, the phase space restrictions
affect the dijet contribution below about 10$\gev^2$, completely
removing the contribution from events with $\ptsq\lsim4\gev^2$.
For $M_X=16.85\gev$, the turn-over is much sharper and the spectrum
cuts off between 1-2$\gev^2$.

\begin{figure}
     \centering
     \subfigure[{\em Donnachie-Landshoff form factor model in
     non-covariant gauge (solid line) and in Feynman gauge
     (solid line). Here
     $\Lambda=0.2\gev^2$ and $\intercept=1.08$.}]{
          \label{fig:dl2858}
		\psfrag{ylabel}{$\frac{1}{\sigma}\frac{\rmd\sigma}{\rmd\ptsq}\gev^2$}
		\psfrag{xlabel}{\small{$\ptsq\gev^2$}}
          \includegraphics[width=.45\textwidth]{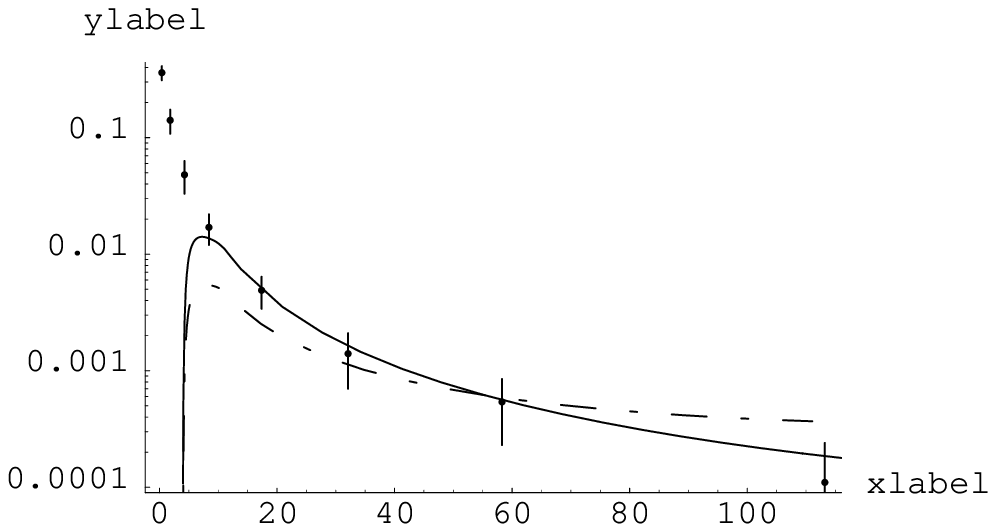}}
     \hspace{.3in}
     \subfigure[{\em Ellis-Ross form factor model in non-covariant
     gauge (solid line) and  Feynman gauge (dashed
     line). Here
     $\Lambda=0.2\gev^2$ and $\intercept=1.08$.}]{
          \label{fig:er2858}
		\psfrag{ylabel}{$\frac{1}{\sigma}\frac{\rmd\sigma}{\rmd\ptsq}\gev^2$}
		\psfrag{xlabel}{\small{$\ptsq\gev^2$}}
          \includegraphics[width=.45\textwidth]{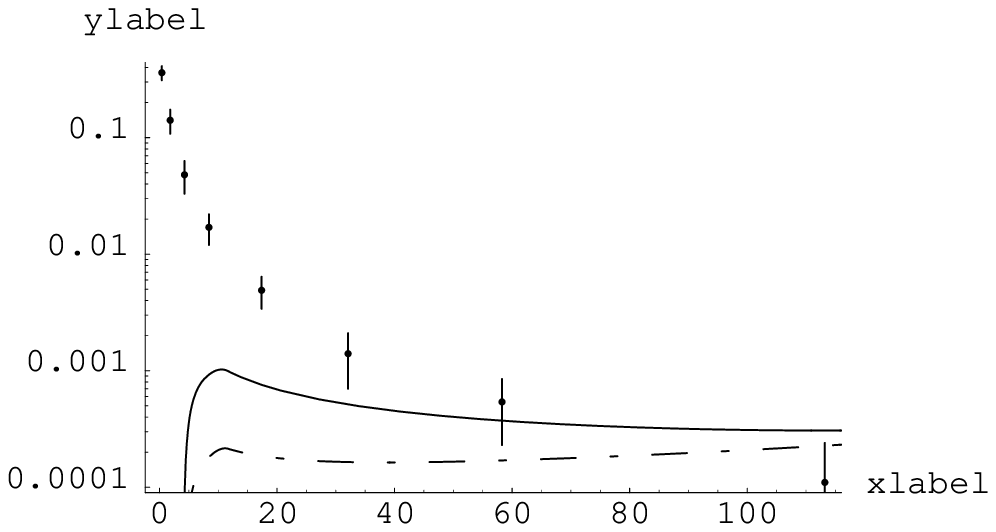}}
     \vspace{.3in}
     \subfigure[{\em Single-gluon exchange model (solid line) and Scalar
     Pomeron exchange model (dashed line).}]{
	   \label{fig:cminusscalar2858}
		\psfrag{ylabel}{$\frac{1}{\sigma}\frac{\rmd\sigma}{\rmd\ptsq}\gev^2$}
		\psfrag{xlabel}{\small{$\ptsq\gev^2$}}
	   \includegraphics[width=.45\textwidth]
		{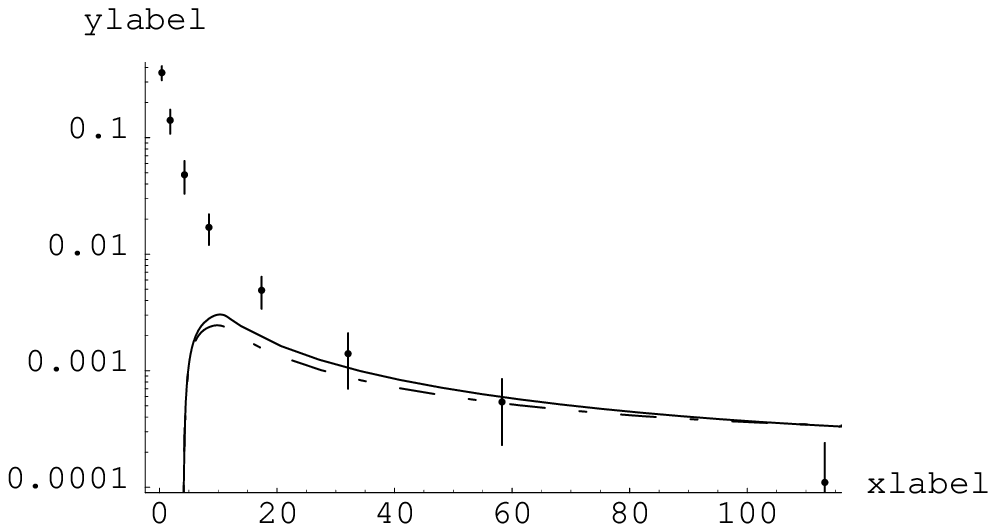}}
     \hspace{.3in}
     \subfigure[{\em Two-gluon exchange model.}]{
	   \label{fig:2glue2858}
		\psfrag{ylabel}{$\frac{1}{\sigma}\frac{\rmd\sigma}{\rmd\ptsq}\gev^2$}
		\psfrag{xlabel}{\small{$\ptsq\gev^2$}}
          \includegraphics[width=.45\textwidth]{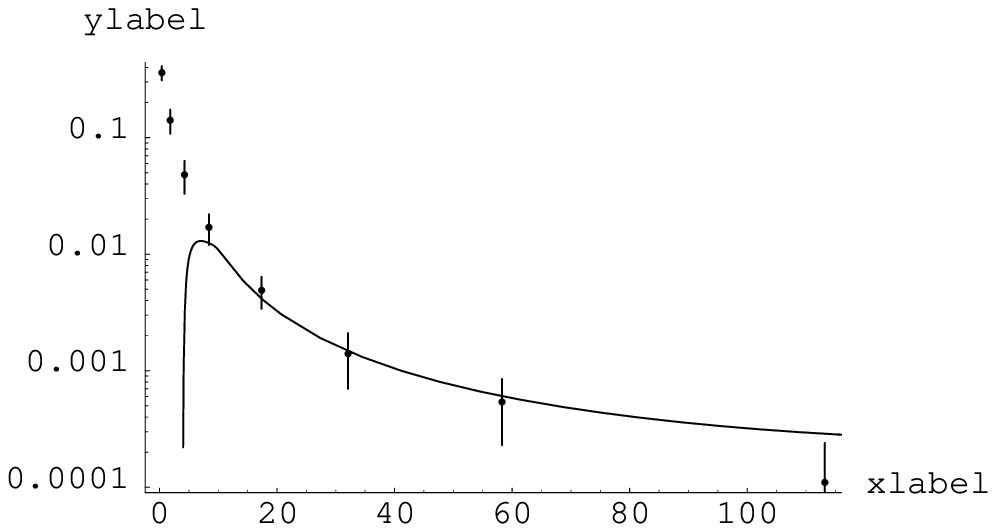}}
     \caption{{\em Fits of colour-singlet exchange models to transverse
     momentum distribution data from\,\cite{Adloff:1997nn}. The fit is
     made to points after the turn-over of the curves (see text for
     explanation). The curves correspond to the models, while the
     solid points are H1 data. The plots show the fits for
     diffractive masses in the region of $M_X=28.58\gev$. The
     statistical and systematic errors have been added in quadrature.}}
     \label{fig:2858multifig}
\end{figure}

\begin{figure}
     \centering
     \subfigure[{\em Donnachie-Landshoff form factor model in
     non-covariant gauge (solid line) and in Feynman gauge
     (dashed line). Here
     $\Lambda=0.2\gev^2$ and $\intercept=1.08$.}]{
          \label{fig:dl1685}
		\psfrag{ylabel}{$\frac{1}{\sigma}\frac{\rmd\sigma}{\rmd\ptsq}\gev^2$}
		\psfrag{xlabel}{\small{$\ptsq\gev^2$}}
          \includegraphics[width=.42\textwidth]{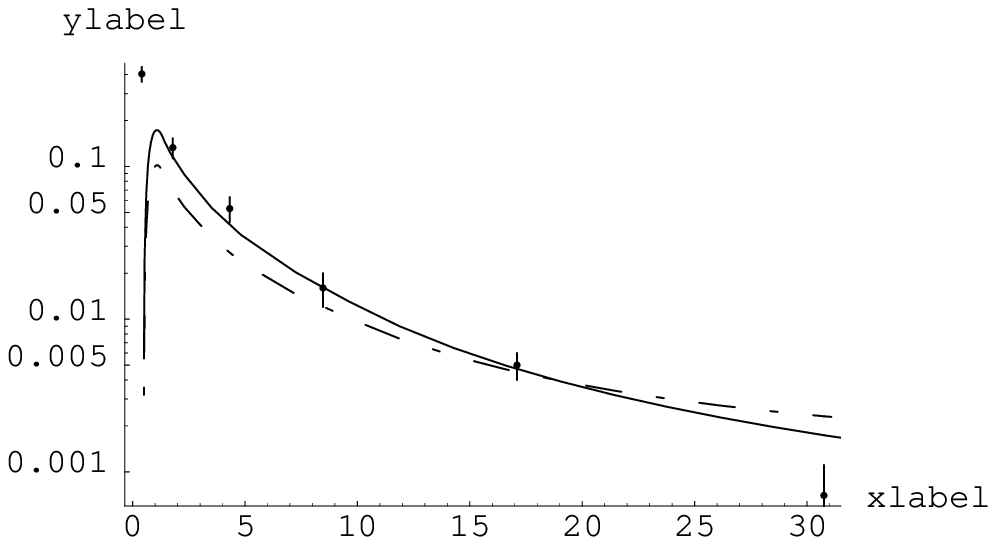}}
	\hspace{.3in}
     \subfigure[{\em Ellis-Ross form factor model in non-covariant
     gauge (solid line) and in Feynman gauge (dashed line). Here
     $\Lambda=0.2\gev^2$ and $\intercept=1.08$.}]{
          \label{fig:er1685}
		\psfrag{ylabel}{$\frac{1}{\sigma}\frac{\rmd\sigma}{\rmd\ptsq}\gev^2$}
		\psfrag{xlabel}{\small{$\ptsq\gev^2$}}
          \includegraphics[width=.42\textwidth]{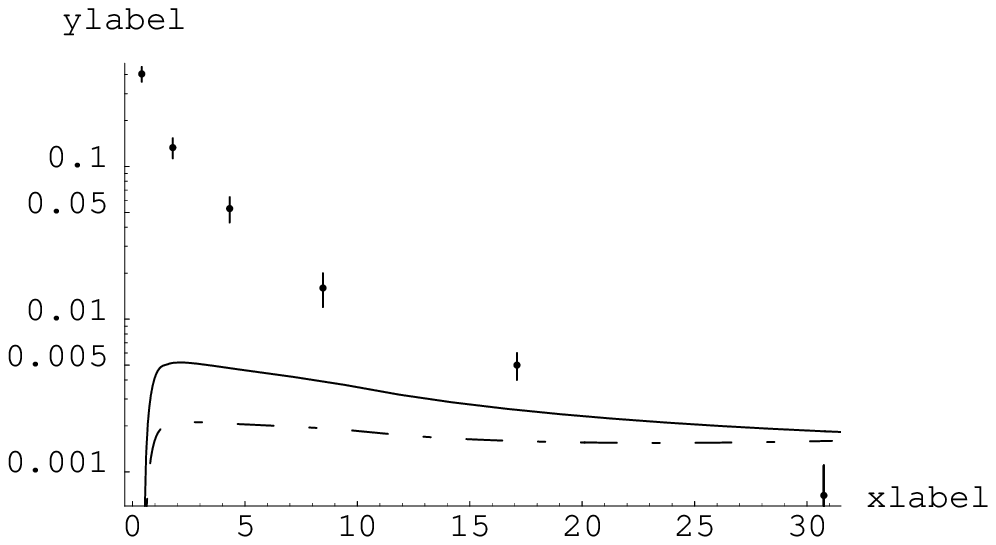}}
     \vspace{.3in}
     \subfigure[{\em Single-gluon exchange model (solid line) and Scalar
     Pomeron exchange model (dashed line).}]{
	   \label{fig:cminusscalar1685}
		\psfrag{ylabel}{$\frac{1}{\sigma}\frac{\rmd\sigma}{\rmd\ptsq}\gev^2$}
		\psfrag{xlabel}{\small{$\ptsq\gev^2$}}
	   \includegraphics[width=.42\textwidth]
		{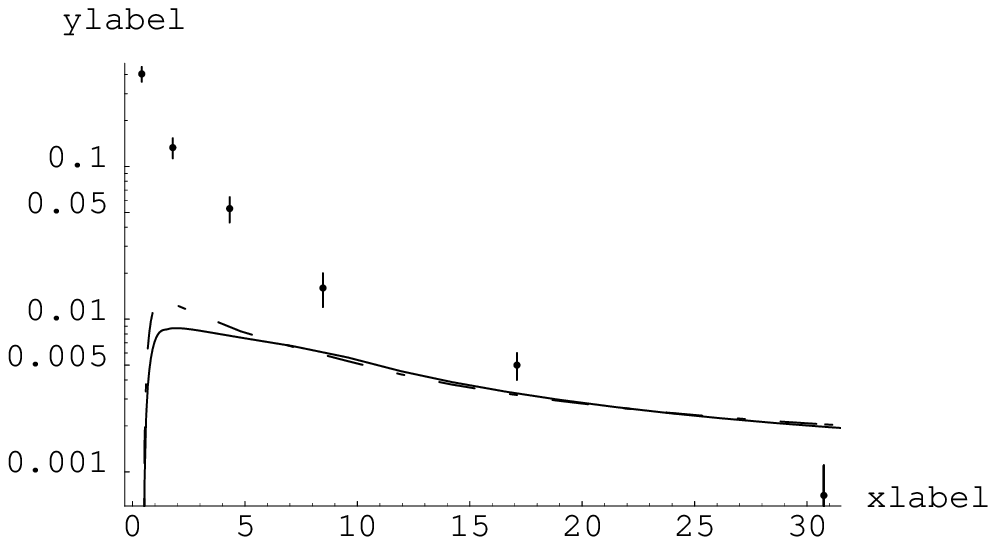}}
\hspace{.3in}
     \subfigure[{\em Two-gluon exchange model.}]{
	   \label{fig:2glue1685}
		\psfrag{ylabel}{$\frac{1}{\sigma}\frac{\rmd\sigma}{\rmd\ptsq}\gev^2$}
		\psfrag{xlabel}{\small{$\ptsq\gev^2$}}
          \includegraphics[width=.42\textwidth]{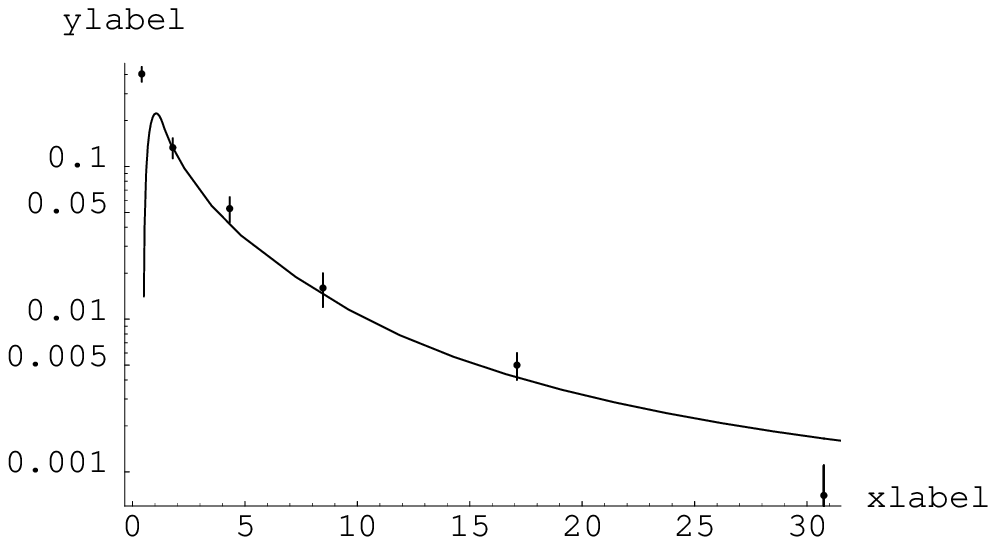}}
     \caption{{\em Fits of colour-singlet exchange models to
     transverse momentum distribution data
     from\,\cite{Adloff:1997nn}. The fit is made to points after the
     turn-over of the curves. The curves correspond to the models,
     while the solid points are H1 data. The plots show the fits for
     diffractive masses in the region of $M_X=16.85\gev$.}}
     \label{fig:1685multifig}
\end{figure}

The curves in Figs.\,\ref{fig:2858multifig} and \ref{fig:1685multifig}
represent the dijet model fits to the H1 data by fitting only the part
of the curves after the phase space-induced turn-over. This shows how
the models describe the large-$\ptsq$ behaviour of the cross
section. In each graph the H1 data\,\cite{Adloff:1997nn} is shown as
solid points with error bars representing statistical and systematic
errors added in quadrature.

The fits shown in Figs.\,\ref{fig:dl2858} and \ref{fig:dl1685} are of
the Donnachie-Landshoff form factor model in the Feynman gauge (solid
line) and in the non-covariant gauge (dashed line). In both cases we
have assumed a pomeron intercept of 1.2 and a low-momentum cut-off in
the form factor of $\Lambda=1.2\gev^2$. These graphs show that the
Donnachie-Landshoff model in the non-covariant gauge provides a good
fit to the large-$\pt$ part of the transverse momentum spectrum, that
is, it predicts a $1/p_{\bot}^4$ fall-off for
$\ptsq\gsim10\gev^2$. The Feynman gauge curve gives a poor fit, mainly
due to a large-$\pt$ tail which is not present in the non-covariant
gauge approach\,\cite{Diehl:1998pd}.

Figs.\,\ref{fig:er2858} and \ref{fig:er1685} show fits of the
Ellis-Ross form factor model to the transverse momentum distribution
data with $\intercept=1.2$ and $\Lambda=1.2\gev^2$. Again, a
high-$\pt$ tail is evident in the cross section calculated in the
Feynman gauge. The E-R model fails to describe the large transverse
momentum behaviour of the data in either gauge, predicting a $1/\ptsq$
fall-off at large transverse momenta. The scalar pomeron and
single-gluon exchange models are shown in
Figs.\,\ref{fig:cminusscalar2858} and
\ref{fig:cminusscalar1685}. These models also fail to describe the
$p_{\bot}^{-4}$ dependence of the data at large $\ptsq$, giving only a
$p_{\bot}^{-2}$ fall-off.

Finally, in Figs.\,\ref{fig:2glue2858} and \ref{fig:2glue1685} we
study the two-gluon model of Diehl\,\cite{Diehl:1995wz}. This model
fits the data well above the phase space-induced low-$\pt$
cut-off. The fit here is similar to the fit obtained using the
Donnachie-Landshoff form factor model.

\section{Comparison of Structure Function and Thrust Transverse
     Momentum Distribution Analyses}\label{sect:3jets}

The results of fitting the colour-singlet exchange models to the
thrust transverse momentum distribution data are surprising when
compared with the results from the comparison of these models with the
large pseudo-rapidity gap structure function data
in\,\cite{Ellis:1998qt}. The structure function data is obtained from
the $\pt$-integrated cross section. The Ellis-Ross and single-gluon
exchange models provided an acceptable fit to the large
pseudo-rapidity gap structure function data, but did were unable to
describe the shape of the thrust transverse momentum distribution
data. By contrast, the Donnachie-Landshoff and two-gluon models fit
the thrust transverse momentum distribution data well, but
under-estimated the structure function data badly at low
$\beta$. There are two possible ways to reconcile these results. The
first possibility is that E-R and single-gluon exchange models may
describe the data at large $\beta$ (in the region where the operator
product expansion is believed to be reliable in describing diffractive
DIS), but that the leading-twist description fails at small
$\beta$. The other possibility is that the D-L and two-gluon models
are able to fit both sets of data with a substantial contribution from
multi-jet diffractive final states. This latter possibility requires
the multi-jet contribution to the low-momentum $\ptsq$ data to be most
significant at small-$\beta$ and to have a thrust transverse momentum
distribution which is strongly peaked near zero.

We discuss these possibilities in more detail in this section.

\subsection{$\beta$-Dependent Structure Function Models}

It is possible that the Ellis-Ross or leading-twist single-gluon
exchange models describe a physical picture of the data for large
$\beta$, but that some other (possibly higher-twist) component is
required at low $\beta$. The E-R model provided a good fit
($\chi^2/{\rm dof}=37/41$) to the large pseudo-rapidity gap structure
function data studied in\,\cite{Ellis:1998qt} and described the data
well over a wide range of $\qsq$, $\beta$ and $\xpom$. Most of the
large pseudo-rapidity gap structure function data is at low $\beta$,
and the transverse momentum distribution data is almost all at low
$\beta$, however, and there is not sufficient large-$\beta$ data
available to test this possibility.

The present thrust transverse momentum distribution analysis
demonstrates a need for a substantial extra non-dijet contribution to
the cross section. Due to the low-$\pt$ dijet cut-off, some additional
contribution is needed to explain the low transverse momentum
data. The position of the cut-off shows also that the size of this
extra contribution is $\beta$-dependent, and peaked at low $\beta$. If
this low-$\beta$ contribution is included in the E-R model fits to
large pseudo-rapidity gap structure function data it will cause the
E-R model to overshoot the structure function data at low $\beta$. The
need for a substantial low-$\pt$, low-$\beta$ contribution to explain
the thrust transverse momentum distribution data therefore presents a
major challenge to E-R and leading-twist models.

\subsection{Two-Gluon Dijet Contribution Plus 3-Jet Contribution}

The $M_X$-dependent position of the dijet low transverse momentum
cut-off is such that at large diffractive masses the region where
dijet events cannot contribute to the diffractive scattering cross
section is larger than for smaller diffractive masses. This dominance
at small $\beta$ is what is expected from quark--antiquark--gluon
diffractive final
states\,\cite{Wusthoff:1997fz,Bartels:1998ea,Bartels:1999tn}. It seems
reasonable therefore, given also the very small values of $\beta$ in
the H1 thrust study, to conclude that the extra contribution required
to explain the thrust transverse momentum distribution spectrum is
probably provided by 3-jet diffractive final-state events.  This
conclusion is consistent with\,\cite{Bartels:1998ea}, where it was
estimated that the 3-jet contribution dominates the diffractive
structure function for $\beta\lsim0.2$, and provides a significant
contribution to $F_2^{D(3)}$ for $\beta\lsim0.4$. This hypothesis is also
consistent with the H1 findings that the average thrust of diffractive
DIS events is less than 0.9\,\cite{Adloff:1997nn}.

If the 3-jet contribution is indeed dominant at small $\beta$ (large
$M_X$), one would expect to see the pattern of diffractive structure
function results found in\,\cite{Ellis:1998qt} for the D-L and
two-gluon models. In this case the dijet contribution is expected to
under-estimate the diffractive structure function at small $\beta$.

Here we discuss the low-$\beta$ component needed to give a good fit to
the thrust transverse momentum distribution data. We assume in this
discussion that the additional contribution at low-$\beta$ originates
from quark--antiquark--gluon diffractive final states.  Considering
the results shown in Figs.\,\ref{fig:2858multifig} and
\ref{fig:1685multifig}, and the analysis in\,\cite{Ellis:1998qt}, one
can immediately make two hypotheses about the distribution of
``3-jet'' processes in the diffractive sample:

\begin{itemize}
\item {\em 3-jet events dominate the large pseudo-rapidity gap
sample at low $\beta$}:
\end{itemize}

The requirement of a large pseudo-rapidity gap between the proton
direction and the diffractive final state removes the low-$\ptsq$
dijet events from the sample\cite{Ellis:1998qt,Williams:1999ze}, hence
higher-multiplicity final states are required to explain the low
transverse momentum part of the spectrum. Furthermore, comparing
Figs.\,\ref{fig:2858multifig} and \ref{fig:1685multifig}, we see also
that the cut-off in the dijet contribution from the phase space
restrictions is shifted to smaller $\ptsq$ for smaller values of
$M_X$. This is to be expected as smaller $M_X$ corresponds to large
values of $\beta$, for which there is a weaker restriction on the
dijet phase space\,\cite{Ellis:1998qt,Williams:1999ze}. Also the ratio
of multi-jet events to dijet events is expected to be smaller at large
$\beta$\,\cite{Bartels:1998ea}.

\begin{itemize}
\item {\em The thrust transverse momentum distribution of 3-jet
events is strongly peaked about 0}:
\end{itemize}

This conclusion is partly based on the success of the two-gluon dijet
model in fitting the $p_{\bot}^{-4}$ behaviour of the cross section at
large transverse momenta, and also on the inability of all the dijet
models to describe the low-$\ptsq$ part of the spectrum due to data
selection cuts. If the 3-jet events were to describe the part of the
transverse momentum spectrum which is not described by dijet events,
one would also require them to have thrust transverse momentum
strongly peaked about $\ptsq=0$, in order to not effect the spectrum
at larger transverse momenta.

A strongly peaked {\em thrust} transverse momentum distribution might
be expected from a two-gluon ``partonic'' model of the pomeron. In
such a model, the leading-order process is one where the photon
interacts with one of the gluons in the pomeron, producing a
diffractive quark--antiquark pair. The third final-state particle is
the other constituent gluon in the pomeron, the pomeron remnant, which
would be traveling in approximately the same direction as the initial
proton. Therefore, especially if there is a reasonably isotropic
distribution of momentum between the two constituent gluons in the
pomeron, one would expect the thrust axis of the final state to be
rather closely aligned with the proton--photon axis. This hypothesis
does, however, require a $\beta$-dependent non-perturbative spreading
of the transverse momentum distribution of the remnant gluon of up to
$\ptsq=6\gev^2$. The intrinsic transverse momentum of the pomeron has
been discussed recently in\,\cite{Nikolaev:1998ys}. It is also
possible that the peaked thrust transverse momentum distribution might
be due to the non-zero momentum transfer at the proton
vertex\,\cite{Breitweg:1998aa}. It is not possible to explore this
latter possibility further at present as there is insufficient data
describing the $t_{\pom}$-dependence of diffractive DIS.

\subsubsection{Parameterization of the Extra Thrust Contribution}

We fit the thrust transverse momentum distribution data with a
combination of the two-gluon model and a parameterization for the
``3-jet'' contribution to diffractive DIS in which the invariant
amplitude of the thrust transverse momentum distribution is described
by the Gaussian\footnote{Note that since the definition of the
transverse momentum variable in this analysis is different to that
used in the diffractive structure function study
in\,\cite{Ellis:1998qt}, the parameterization used to fit the thrust
transverse momentum distributions here cannot be directly applied to
the structure function analysis.}:

\begin{equation}
f(Q^2,\beta)=\frac{1}{\qsq\beta}e^{-\frac{\ptsq}{\beta^2\ptsqmax}}.
\end{equation}

\noindent
This provides a $\beta$-dependent distribution which falls sharply
with increasing $\ptsq$. Since the H1 data contains no
information about the absolute normalisation, in these fits the
overall normalisation of each contribution is a free parameter, giving
two adjustable parameters when combining the two contributions. The
most unsatisfactory feature of the ``3-jet'' interpretation of this
extra contribution is the need for an exponential $\beta$-dependent
cut-off in the thrust transverse momentum distribution. 

\begin{figure}
     \centering
     \subfigure[{\em $M_X=28.58\gev$}]{
          \label{fig:3jets2858}
		\psfrag{ylabel}{$\frac{1}{\sigma}\frac{\rmd\sigma}{\rmd\ptsq}\gev^2$}
		\psfrag{xlabel}{\small{$\ptsq\gev^2$}}
          \includegraphics[width=.45\textwidth]{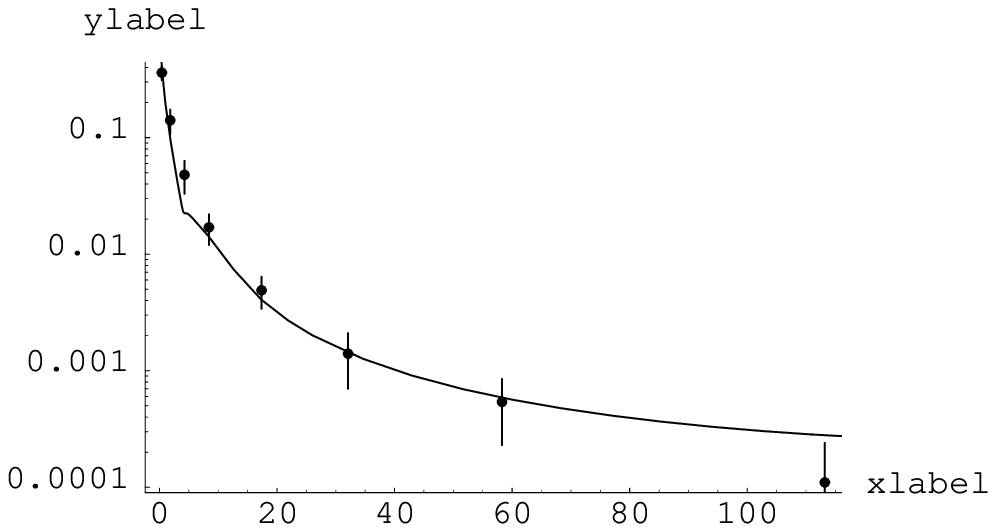}}
     \hspace{.3in}
     \subfigure[{\em $M_X=21.20\gev$}]{
          \label{fig:3jets2120}
		\psfrag{ylabel}{$\frac{1}{\sigma}\frac{\rmd\sigma}{\rmd\ptsq}\gev^2$}
		\psfrag{xlabel}{\small{$\ptsq\gev^2$}}
          \includegraphics[width=.45\textwidth]{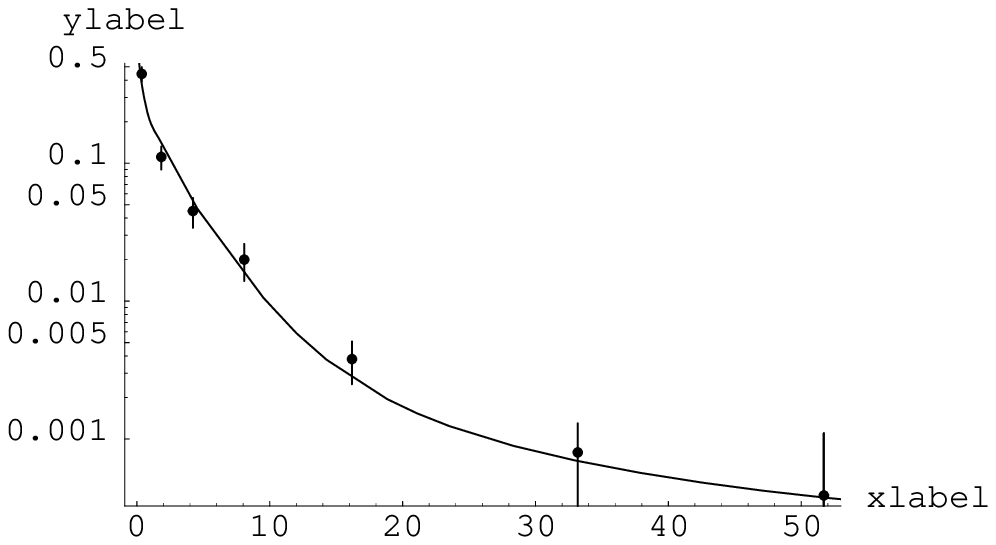}}\\
     \vspace{.3in}
     \subfigure[{\em $M_X=16.85\gev$}]{
	   \label{fig:3jets1685}
		\psfrag{ylabel}{$\frac{1}{\sigma}\frac{\rmd\sigma}{\rmd\ptsq}\gev^2$}
		\psfrag{xlabel}{\small{$\ptsq\gev^2$}}
	   \includegraphics[width=.45\textwidth]
		{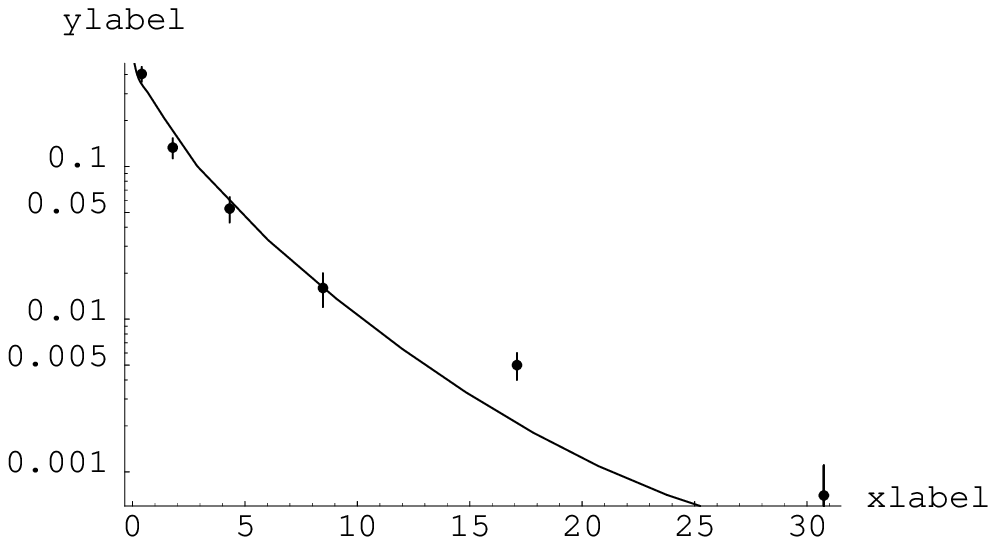}}
	\hspace{.3in}
     \subfigure[{\em $M_X=12.82\gev$}]{
	   \label{fig:3jets1282}
		\psfrag{ylabel}{$\frac{1}{\sigma}\frac{\rmd\sigma}{\rmd\ptsq}\gev^2$}
		\psfrag{xlabel}{\small{$\ptsq\gev^2$}}
          \includegraphics[width=.45\textwidth]{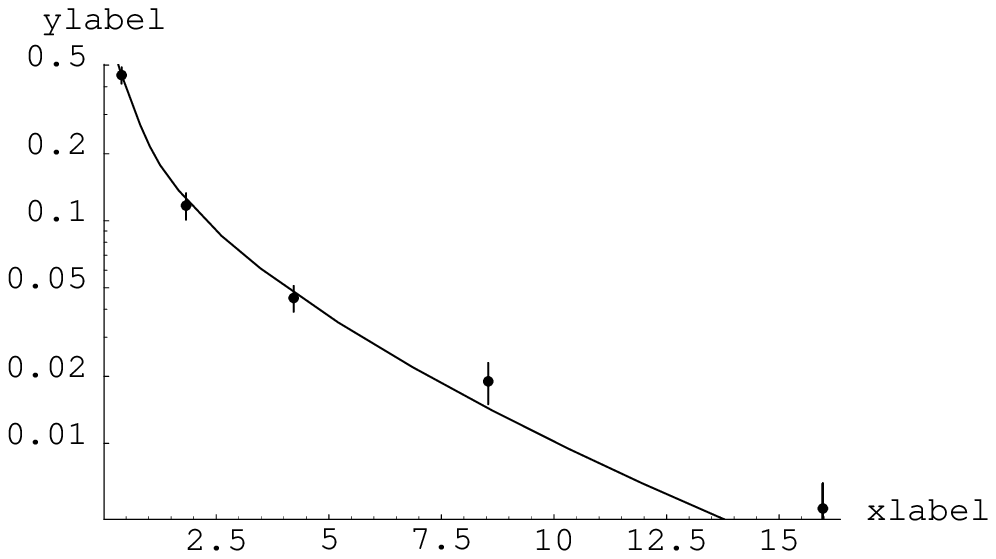}}\\
     \vspace{.3in}
     \subfigure[{\em $M_X=9.40\gev$}]{
	   \label{fig:3jets940}
		\psfrag{ylabel}{$\frac{1}{\sigma}\frac{\rmd\sigma}{\rmd\ptsq}\gev^2$}
		\psfrag{xlabel}{\small{$\ptsq\gev^2$}}
	   \includegraphics[width=.45\textwidth]
		{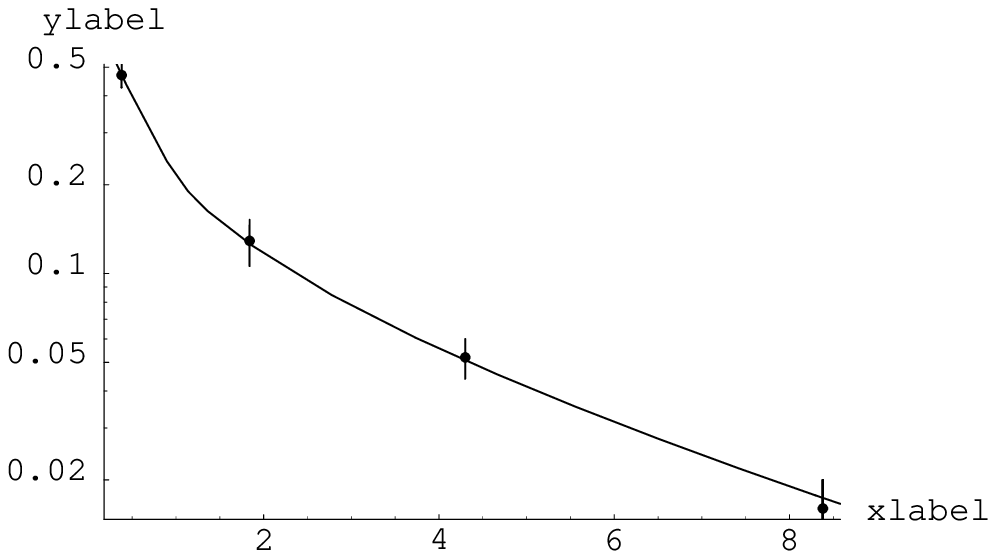}}
	\hspace{.3in}
     \subfigure[{\em $M_X=7.01\gev$}]{
	   \label{fig:3jets701}
		\psfrag{ylabel}{$\frac{1}{\sigma}\frac{\rmd\sigma}{\rmd\ptsq}\gev^2$}
		\psfrag{xlabel}{\small{$\ptsq\gev^2$}}
          \includegraphics[width=.45\textwidth]{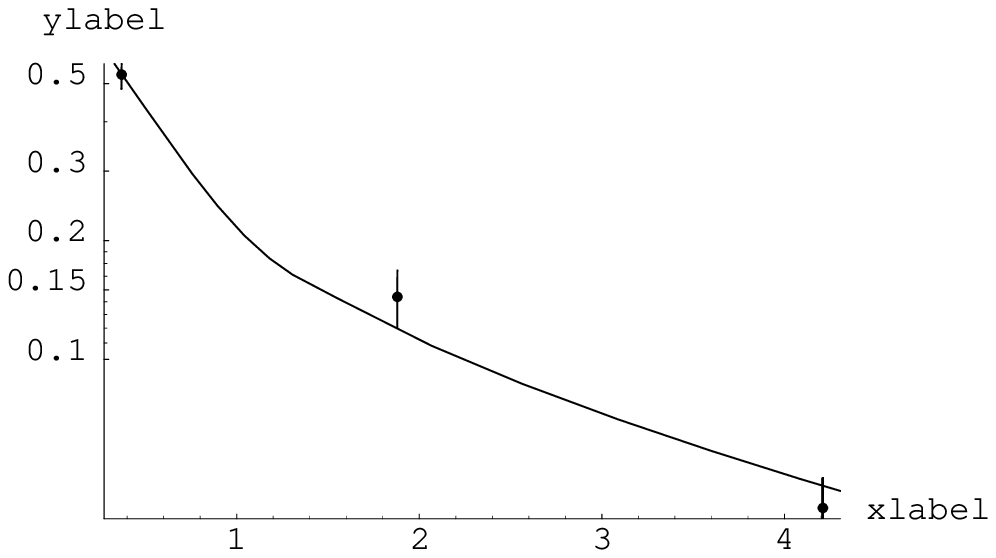}}
     \caption{{\em Comparison of the combination of a two-gluon dijet
	   model and a 3-jet parameterization with thrust transverse momentum
	   distribution data from\,\cite{Adloff:1997nn} at different
	   diffractive masses.}}
     \label{fig:3jets2gluemultifig}
\end{figure}

The combination fit is shown in Fig.\,\ref{fig:3jets2gluemultifig}.
This fit describes the transverse momentum distribution data well in
the range of diffractive masses studied by H1, only underestimating
the data slightly at $M_X=16.85\gev$. This confirms that the data may
be described by the combination of a dijet model with a quadratic
fall-off with transverse momentum and a ``multi-jet'' parameterization
with an exponentially decaying thrust transverse momentum
distribution.

A more complete test of this hypothesis would be to study data for
which the overall normalisation of the cross section at each
diffractive mass is measured, as this would enable allow the relative
contributions of 2- and 3-jet final states to the diffractive cross
section to be measured.

\section{Discussion and Conclusions} \label{sect:conclusions}

We have studied the transverse momentum distribution of diffractive
DIS and compared the dijet component of several diffractive DIS models
with recent H1 data\,\cite{Adloff:1997nn}.

The general trend in the data is towards a $1/\ptsq$ fall-off at low
transverse momenta, steepening to $1/p_{\bot}^4$ at about
4-10$\gev^2$. The position of the change in slope depends on the
diffractive mass, with the change in slope occurring at larger values
of the transverse momentum variable for larger $M_X$.

The data selection cuts, particularly the pseudo-rapidity cut imposed
by H1 to select diffractive events, lead to a restriction on the
available phase space for a wide of the parameter space covered by
HERA. The main result is that, for most diffractive masses and for the
range of $\qsq$ and $\xpom$ in this analysis, data selection cuts are
expected to remove all dijet events with small transverse momenta. As
a result, one requires a further contribution, for example
quark--antiquark--gluon final states, to explain the small-$\ptsq$
region of the H1 transverse momentum distribution data. The region in
$\pt$ where the data changes slope corresponds rather closely to the
region where phase space effects due to data selection cuts remove the
low-$\pt$ dijet contribution.

We compared four classes of colour-singlet exchange models with the H1
data: two form factor vector pomeron models, a leading-twist
single-gluon exchange model, a scalar pomeron model, and a two-gluon
exchange model. The cross section from the form factor models was
calculated in both the Feynman gauge and in a non-covariant gauge. The
difference between the two gauge choices was particularly evident in
the large-$\ptsq$ tail which one obtains using the Feynman gauge,
which is not seen in any other approach, and which fails to describe
the present data. The Donnachie-Landshoff form factor model (in
non-covariant gauge) provided a good fit to the large transverse
momentum part of the spectrum for all values of diffractive mass
beyond the point where the slope changed to a $1/p_{\bot}^4$ fall. The
Ellis-Ross form factor model, as well as the single-gluon and scalar
pomeron models, predicts a $1/\ptsq$ fall at large transverse momenta
which is not consistent with the H1 data. The two-gluon model fit the
large-$\pt$ part of the transverse momentum distribution spectrum
well.

Since the region where dijet events cannot contribute to the
diffractive cross section is much larger for larger diffractive masses
(i.e. smaller $\beta$), it was concluded that the extra contribution
required to describe the H1 transverse momentum distribution data was
probably from 3-jet diffractive final states. It was found that a
combination of the two-gluon dijet model and a simple parameterization
for the extra contribution using $\beta$-dependent Gaussian gave a
good fit to the entire thrust transverse momentum distribution
spectrum.

The 3-jet interpretation of this extra contribution is consistent with
the earlier structure function analysis of large pseudo-rapidity gap
diffractive DIS\,\cite{Ellis:1998qt}. In this interpretation a large
3-jet diffractive contribution is required at small $\beta$ in
addition to the dijet contribution from the Donnachie-Landshoff or
two-gluon models. In addition, however, the present analysis also
requires the 3-jet diffractive events to have thrust transverse
momentum strongly peaked about zero. This analysis also suggests that
the apparent success of the Ellis-Ross, single-gluon and scalar
pomeron models in describing the diffractive structure function may be
due to these models over-estimating the large-$\ptsq$ dijet
contribution.

An alternative possibility is that the E-R or leading-twist
single-gluon models may provide a reasonable fit to the diffractive
DIS cross section at moderate to large values of $\beta$. Present
pseudo-rapidity gap thrust transverse momentum distribution and
structure function data in which the requirement of a large
pseudo-rapidity gap places strong restrictions on the available phase
space corresponds mostly to small $\beta$ and at present there is
insufficient data, particularly at large-$\beta$, to either prove or
disprove these hypotheses.

This analysis demonstrates the worth of studying the transverse
momentum dependence in distinguishing between models. Clearly it is of
importance to obtain an improved data set in the diffractive
scattering regime. In particular, it would be helpful if smaller
$\xpom$ data were available as the present jet production data has
relatively large $\xpom$ where the purely diffractive interpretation
may be invalid.

Further investigation is required to test the hypotheses presented in
this paper. On the theory side, there is a need for a full 3-jet
diffractive DIS calculation or at least, as far as testing this
analysis is concerned, for a proper Monte Carlo treatment of thrust
transverse momentum distribution of 3-jet diffractive events with
proper account taken of experimental data selection cuts. Better
statistics and finer binning in the thrust transverse momentum
variable would allow one to explore the interface between diffractive
dijet production and the low-$\ptsq$ contribution.  In a similar vein,
analysing data corresponding to larger values of $\beta$, where phase
space effects due to pseudo-rapidity cuts are less significant and
dijet events are expected to dominate over higher-multiplicity
diffractive final states, one would expect to see the quartic
$\pt$-dependence in the present H1 data extend to smaller values of
the $\pt$. On the other hand, restricting to smaller values of
$\beta$, one would be able to push the position of the change to a
quartic fall-off to larger values of $\ptsq$.

A more interesting experiment would be to keep the same range of
kinematic parameters in $\qsq$, $\beta$ and $\xpom$ and impose a
strongly pseudo-rapidity cut. This would push the position of the
low-momentum dijet cut-off to larger values of $\ptsq$. The only
effect on the sharply-peaked low thrust-$\pt$ contribution would be a
decrease in the overall normalisation. With fine enough binning in
$\pt$ this should therefore lead to the production of a ``dip'' in the
cross section in the $\ptsq$ region between where the exponential
low-$\pt$ contribution cuts off and where the $1/p_{\bot}^4$ dijet
contribution starts.

Information about the overall normalisation of the cross section would
enable the relative contributions of dijet and 3-jet events to
diffractive DIS to be measured.

\section*{Acknowledgments}

Thanks to my collaborators, Graham Ross and John Ellis, and to Markus
Diehl for helpful discussions. Thanks also to Ian Plummer for reading
the manuscript and for very helpful comments. I am also very grateful
for financial support from the NZFUW through an NZFUW Post Graduate
Fellowship and a Sadie Balkind Scholarship.

\section*{Appendix.~~Structure Function Analysis of Form Factor Models in
Non-Covariant Gauge}

This paper extends the form factor analysis of\,\cite{Ellis:1998qt} to
the non-covariant gauge choice described in\,\cite{Diehl:1998pd}. It
is interesting to see whether the form factor models describe the
large pseudo-rapidity gap structure function data in this gauge. We
have also repeated the structure function analysis for the form factor
models with a low-momentum cut-off of $\Lambda=1.2\gev^2$ in the form
factor, as suggested by fits to vector meson production data by
Donnachie and Landshoff\,\cite{Donnachie:1987pu}, rather than the
smaller value of $\Lambda=0.2\gev^2$ used in the original
pseudo-rapidity gap structure function
analysis\,\cite{Ellis:1998qt}. We compared the Donnachie-Landshoff and
Ellis-Ross form factor models with the large virtuality constraint
data from\,\cite{Phillips:1995jpp}. Fitting to the 42 points
considered in\,\cite{Ellis:1998qt}, we obtained the $\chi^2$
parameters shown in Table\,\ref{table:newchisq}.

\begin{table}[htp]
\centering
\begin{tabular}{|c|cc|}
\hline
&  &  \\ [-8pt]
& \multicolumn{2}{c|}{$\chi^2/{\rm dof}$} \\ 
&  &  \\ [-8pt]\cline{2-3}
&  &  \\ [-8pt]
& ~~$\alpha_{\pom}(0)=1.08$~~ & ~~$\alpha_{\pom}(0)=1.2$~~ \\
&  &  \\[-8pt] \hline
&  &  \\[-8pt]
E-R: $f(k^2)=\mbox{\Large $\sqrt{\frac{\Lambda^2}{\Lambda^2-k^2}}$}$ &
39/41~(37/41) & 57/41 (54/41) \\  
&  &  \\ [-8pt]
&  &  \\ [-8pt]
D-L: $f(k^2)=\mbox{\Large $\frac{\Lambda^2}{\Lambda^2-k^2}$}$ &
81/41~(102/41) &  79/41 (97/41) \\ 
&  &  \\ [-8pt]
&  &  \\ [-8pt]
E-R: (non-covariant gauge) &  41/41 (41/41)       &  56/41 (55/41)      \\ 
&  &  \\ [-8pt]
&  &  \\ [-8pt]
D-L: (non-covariant gauge) &  69/41 (86/41)         & 68/41 (82/41)       \\ 
&  &  \\ [-2pt]\hline
\end{tabular}
\caption[Result of the fit of the various colour-singlet exchange
models to~$F_{2}^{D(3)}$ structure function data.]{{\it Result of the
fit of form factor models to large pseudo-rapidity gap diffractive
structure function data from\,\cite{Phillips:1995jpp}. The fits assume
a cut-off in the form factors of $\Lambda^2=1.2\gev^2$. Results of
fits with a form factor cut-off of $\Lambda^2=0.2\gev^2$ in the form
factors, as was used in the original analysis in\,\cite{Ellis:1998qt}
are shown in brackets.}}
\label{table:newchisq}
\end{table}

We can see from Table\,\ref{table:newchisq} that changing the
low-momentum form factor cut-off, $\Lambda$, and calculating the cross
section in the non-covariant gauge make no real difference in the fits
obtained in the Ellis-Ross model. The most significant effect on the
Donnachie-Landshoff fits is from using the larger form factor
cut-off. The D-L model, calculated in the non-covariant gauge with a
form factor cut-off of $\Lambda^2=1.2\gev^2$, gives a moderately
successful fit to the diffractive structure function data. The
principal reason that the results for form factor cut-offs
$\Lambda^2=0.2\gev^2$ or $\Lambda^2=1.2\gev^2$ are similar is that
much of the data we are studying corresponds to an exchanged quark
virtuality of $k^2\gsim$ a few GeV$^2$ in the form factors, hence the
low-momentum cut-off in the form factor has little effect. The
difference in the cross sections between using the Feynman gauge or
the non-covariant gauge is visible when the unintegrated transverse
momentum distributions are studied, as can be seen in the graphs in
Sect.\,\ref{sect:ptsqanalysis}.

\providecommand\singleletter[1]{#1}


\begin{thebibliography}{10}

\bibitem{Ellis:1996cg}
J.~Ellis and G.~G. Ross,
\newblock Phys. Lett. {\bf B384}, 293 (1996).

\bibitem{Donnachie:1987pu}
A.~Donnachie and P.~V. Landshoff,
\newblock Phys. Lett. {\bf 185B}, 403 (1987).

\bibitem{Donnachie:1987xh}
A.~Donnachie and P.~V. Landshoff,
\newblock Phys. Lett. {\bf 191B}, 309 (1987).

\bibitem{Vermaseren:1996iy}
J.~Vermaseren, F.~Barreiro, L.~Labarga, and F.~J. Yndurain,
\newblock Phys. Lett. {\bf B418}, 363 (1998).

\bibitem{Landshoff:1987yj}
P.~V. Landshoff and O.~Nachtmann,
\newblock Z. Phys. {\bf C35}, 405 (1987).

\bibitem{Diehl:1995wz}
M.~Diehl,
\newblock Z. Phys. {\bf C66}, 181 (1995).

\bibitem{Diehl:1998pd}
M.~Diehl,
\newblock Eur. Phys. J. {\bf C6}, 503 (1999).

\bibitem{Bartels:1998ea}
J.~Bartels, J.~Ellis, H.~Kowalski, and M.~Wusthoff,
\newblock Eur. Phys. J. {\bf C7}, 443 (1999).

\bibitem{Ellis:1998qt}
J.~Ellis, G.~G. Ross, and J.~Williams,
\newblock Eur. Phys. J. {\bf C10}, 443 (1999).

\bibitem{Donnachie:1992rh}
A.~Donnachie and P.~V. Landshoff,
\newblock Phys. Lett. {\bf B285}, 172 (1992).

\bibitem{Buchmuller:1997xw}
W.~Buchm{\"{u}}ller, M.~F. McDermott, and A.~Hebecker,
\newblock Nucl. Phys. {\bf B487}, 283 (1997).

\bibitem{Buchmuller:1997eb}
W.~Buchm{\"{u}}ller, M.~F. McDermott, and A.~Hebecker,
\newblock Phys. Lett. {\bf B410}, 304 (1997).

\bibitem{Adloff:1997nn}
H1, C.~Adloff {\em et~al.},
\newblock Eur. Phys. J. {\bf C1}, 495 (1998).

\bibitem{Wusthoff:1997fz}
M.~Wusthoff,
\newblock Phys. Rev. {\bf D56}, 4311 (1997).

\bibitem{Bartels:1999tn}
J.~Bartels, H.~Jung, and M.~Wusthoff,
\newblock hep-ph/9903265.

\bibitem{Williams:1999ze}
J.~Williams,
\newblock hep-ph/9905574.

\bibitem{Donnachie:1984hf}
A.~Donnachie and P.~V. Landshoff,
\newblock Nucl. Phys. {\bf B231}, 189 (1984).

\bibitem{Donnachie:1984xq}
A.~Donnachie and P.~V. Landshoff,
\newblock Nucl. Phys. {\bf B244}, 322 (1984).

\bibitem{Donnachie:1986iz}
A.~Donnachie and P.~V. Landshoff,
\newblock Nucl. Phys. {\bf B267}, 690 (1986).

\bibitem{Donnachie:1992ny}
A.~Donnachie and P.~V. Landshoff,
\newblock Phys. Lett. {\bf B296}, 227 (1992).

\bibitem{Ahmed:1994nw}
H1, T.~Ahmed {\em et~al.},
\newblock Nucl. Phys. {\bf B429}, 477 (1994).

\bibitem{Ahmed:1995ui}
H1, T.~Ahmed {\em et~al.},
\newblock Nucl. Phys. {\bf B435}, 3 (1995).

\bibitem{Derrick:1993xh}
ZEUS, M.~Derrick {\em et~al.},
\newblock Phys. Lett. {\bf B315}, 481 (1993).

\bibitem{Nikolaev:1998ys}
N.~N. Nikolaev,
\newblock hep-ph/9905562.

\bibitem{Breitweg:1998aa}
ZEUS, J.~Breitweg {\em et~al.},
\newblock Eur. Phys. J. {\bf C1}, 81 (1998).

\bibitem{Phillips:1995jpp}
J.~P. Phillips,
\newblock {\em The Deep-Inelastic Structure of Diffraction},
\newblock PhD thesis, University of Manchester, 1995,
\newblock available from\hfill\\ {\tt
  http://www-h1.desy.de/h1/www/h1work/dif/publications.html}.

\end{thebibliography}
\end{document}